# On the Nature of the Change in the Wave Function in a Measurement in Quantum Mechanics

Douglas M. Snyder

## Abstract

Generally a central role has been assigned to an unavoidable physical interaction between the measuring instrument and the physical entity measured in the change in the wave function that often occurs in measurement in quantum mechanics. A survey of textbooks on quantum mechanics by authors such as Dicke and Witke (1960), Eisberg and Resnick (1985), Gasiorowicz (1974), Goswami (1992), and Liboff (1993) supports this point. Furthermore, in line with the view of Bohr and Feynman, generally the unavoidable interaction between a measuring instrument and the physical entity measured is considered responsible for the uncertainty principle. A gedankenexperiment using Feynman's double-hole interference scenario shows that physical interaction is not necessary to effect the change in the wave function that occurs in measurement in quantum mechanics. Instead, the general case is that knowledge is linked to the change in the wave function, not a physical interaction between the physical existent measured and the measuring instrument. Empirical work on electron shelving that involves null measurements, or what Renninger (1960) called negative observations, supports these points. Work on electron shelving is reported by Dehmelt and his colleagues (1986), Wineland and his colleagues (1986), and Sauter, Neuhauser, Blatt, and Toschek (1986).

## Text

Generally the change in the wave function that often occurs in measurement in quantum mechanics has been ascribed to the unavoidable physical interaction between the measuring instrument and the physical entity measured. Indeed, Bohr (1935) maintained that this unavoidable interaction was responsible for the uncertainty principle, more specifically the inability to simultaneously measure observable quantities described by non-commuting Hermitian operators (e.g., the position and momentum of a particle). The following series of gedankenexperiments in this section will show that this interaction is not necessary to effect a change in the wave function. The series of gedankenexperiments indicates that knowledge plays a significant role in the change in the wave function that often occurs in measurement (Snyder, 1996a, 1996b).

# On the Nature

GEDANKENEXPERIMENT 1

Feynman, Leighton, and Sands (1965) explained that the distribution of electrons passing through a wall with two suitably arranged holes to a backstop where the positions of the electrons are detected exhibits interference (Figure 1). Electrons at the backstop may be detected with a Geiger counter or an electron multiplier. Feynman et al. explained that this interference is characteristic of wave phenomena and that the distribution of electrons at the backstop indicates that each of the electrons acts like a wave as it passes through the wall with two holes. It should be noted that when the electrons are detected in this gedankenexperiment, they are detected as discrete entities, a characteristic of particles, or in Feynman et al.'s terminology, "lumps" (p. 1-5).

In Figure 1, the absence of lines indicating possible paths for the electrons to take from the electron source to the backstop is not an oversight. An electron is not taking one or the other of the paths. Instead, the wave function associated with each electron after it passes through the holes is the sum of two more elementary wave functions, with each of these wave functions experiencing diffraction at one or the other of the holes. Epstein (1945) emphasized that when the quantum mechanical wave of some physical entity such as an electron exhibits interference, it is interference generated only in the wave function characterizing the individual entity.

The diffraction patterns resulting from the waves of the electrons passing through the two holes would at different spatial points along a backstop behind the hole exhibit constructive or destructive interference. At some points along the backstop, the waves from each hole sum (i.e., constructively interfere), and at other points along the backstop, the waves from each hole subtract (i.e., destructively interfere). The distribution of electrons at the backstop is given by the absolute square of the combined waves at different locations along the backstop, similar to the characteristic of a classical wave whose intensity at a particular location is proportional to the square of its amplitude. Because the electrons are detected as discrete entities, like particles, at the backstop, it takes many electrons to determine the intensity of the quantum wave that describes each of the electrons and that is reflected in the distribution of the electrons against the backstop.

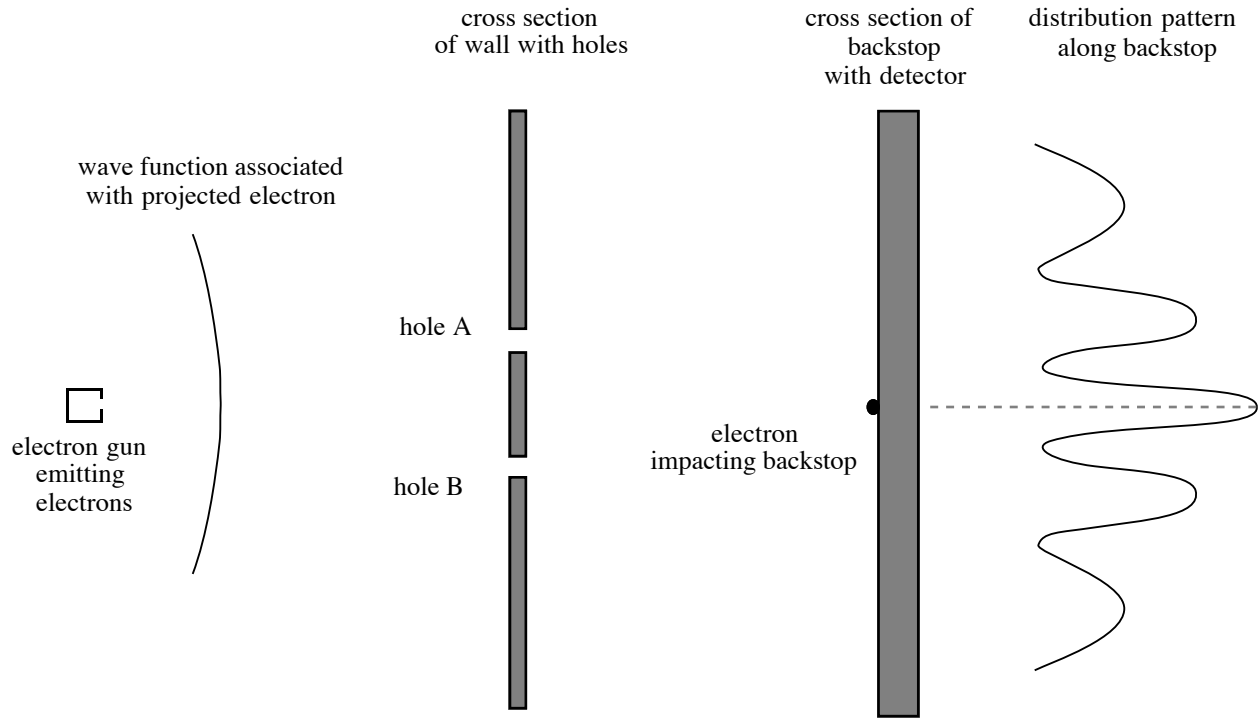

Figure 1

Two-hole gedankenexperiment in which the distribution of electrons reflects interference in the wave functions of electrons. (Gedankenexperiment 1)

# On the Nature

## GEDANKENEXPERIMENT 2

Feynman et al. further explained that if one were to implement a procedure in which it could be determined through which hole the electron passed, the interference pattern is destroyed and the resulting distribution of the electrons resembles that of classical particles passing through the two holes in an important way. Feynman et al. relied on a strong light source behind the wall and between the two holes that illuminates an electron as it travels through either hole (Figure 2). Note the significant difference between the distribution patterns in Figures 1 and 2.

In Figure 2, the path from the electron's detection by the light to the backstop is indicated, but it is important to emphasize that this path is inferred only after the electron has reached the backstop. A measurement of the position of the electron with the use of the light source introduces an uncertainty in its momentum. Only when the electron is detected at the backstop can one infer the path the electron traveled from the hole it went through to the backstop. It is not something one can know before the electron strikes the backstop.

In Feynman et al.'s gedankenexperiment using the light source, the distribution of electrons passing through both holes would be similar to that found if classical particles were sent through an analogous experimental arrangement in an important way. Specifically, as in the case of classical particles, this distribution of electrons at the backstop is the simple summation of the distribution patterns for electrons passing through one or the other of the holes. Figure 3 shows the distribution patterns of electrons passing through hole A and electrons passing through hole B in Gedankenexperiment 2. These distribution patterns are identical to those that would occur if only one or the other of the holes were open at a particular time. An inspection of Figure 3 shows that summing the distribution patterns for the electrons passing through hole A and those passing through hole B results in the overall distribution of electrons found in Gedankenexperiment 2.

## THE UNCERTAINTY PRINCIPLE

Feynman et al.'s gedankenexperiments are themselves very interesting in that they illustrate certain apparently incongruent characteristics of microscopic physical existents, namely particle-like and wave-like features. Feynman et al. discussed their gedankenexperiments in terms of Heisenberg's uncertainty principle. Feynman et al. wrote:

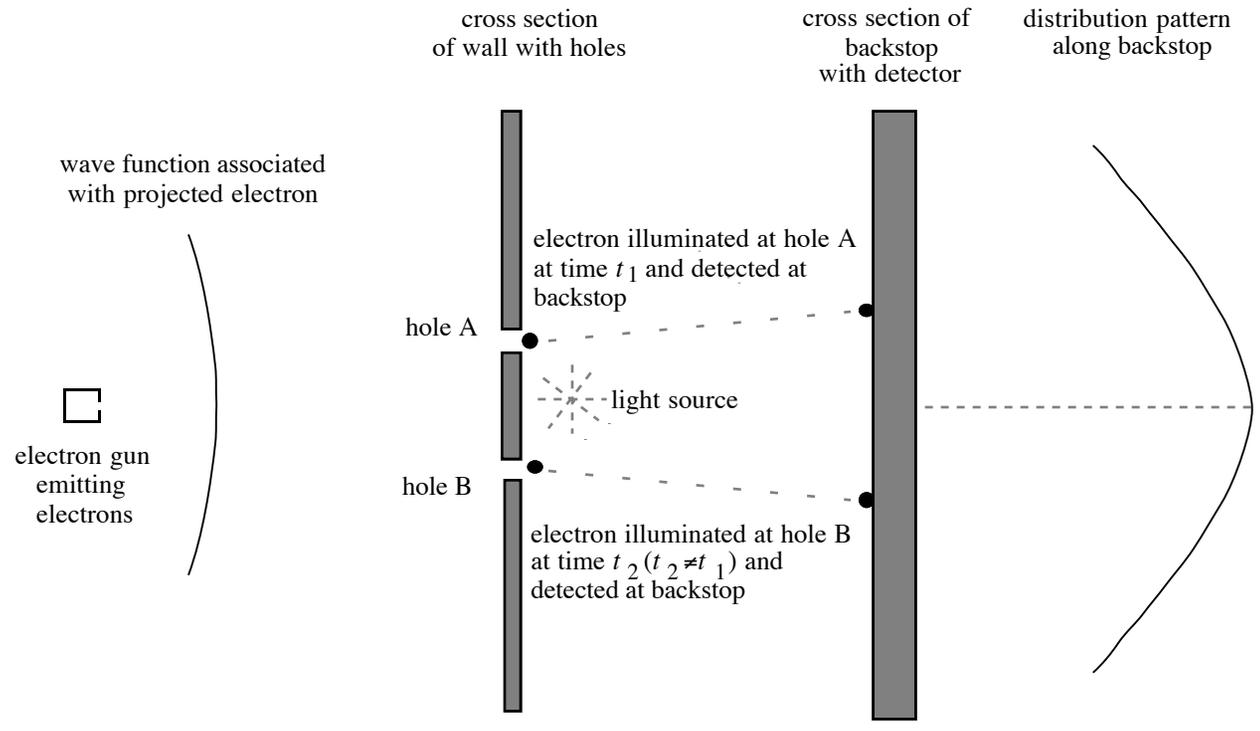

Figure 2
Two-hole gedankenexperiment with strong light source.
(Gedankenexperiment 2)

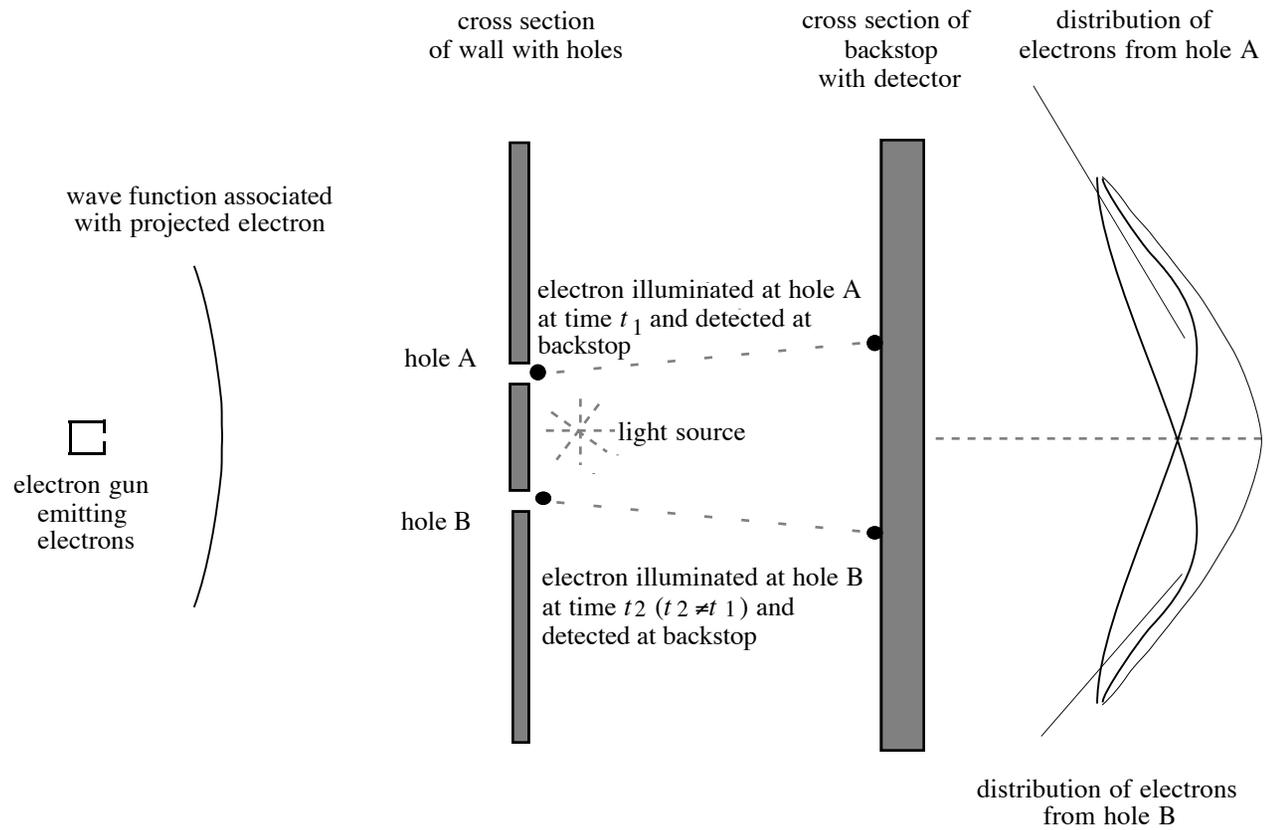

Figure 3

Two-hole gedankenexperiment with strong light source in which the distribution of electrons from each hole is shown.

> He [Heisenberg] proposed as a general principle, his *uncertainty principle*, which we can state in terms of our experiment as follows: "It is impossible to design an apparatus to determine which hole the electron passes through, that will not at the same time disturb the electrons enough to destroy the interference pattern." If an apparatus is capable of determining which hole the electron goes through, it *cannot* be so delicate that it does not disturb the pattern in an essential way. (p. 1-9)

Note that Feynman et al. implied in their description of the uncertainty principle that there is an unavoidable interaction between the measuring instrument (in their gedankenexperiment, the strong light source emitting photons) and the physical entity measured. Feynman et al. also wrote concerning Gedankenexperiment 2:

> the jolt given to the electron when the photon is scattered by it is such as to change the electron's motion enough so that if it might have gone to where $P_{12}$ [the electron distribution] was at a maximum [in Gedankenexperiment 1] it will instead land where $P_{12}$ was at a minimum; that is why we no longer see the wavy interference effects. (p. 1-8)

In determining through which hole an electron passes, Feynman et al., like most physicists, maintained that the electrons are unavoidably disturbed by the photons from the light source and it is this disturbance by the photons that destroys the interference pattern. Indeed, in a survey of a number of the textbooks of quantum mechanics, it is interesting that each author, in line with Feynman and Bohr, allowed a central role in the change in the wave function that occurs in a measurement to a physical interaction between the physical existent measured and some physical measuring apparatus. The authors of these textbooks are Dicke and Witke (1960), Eisberg and Resnick (1974/1985), Gasiorowicz (1974), Goswami (1992), Liboff (1993), Merzbacher (1961/1970), and Messiah (1962/1965).

It is important to note explicitly that some causative factor is necessary to account for the very different distributions of the electrons in Figures 1 and 2. Feynman et al. maintained that the physical interaction between the electrons and photons from the light source is this factor.

GEDANKENEXPERIMENT 3

Feynman et al.'s gedankenexperiments indicate that in quantum mechanics the act of taking a measurement in principle is linked to, and often affects, the physical world which is being measured. The nature of taking a measurement in quantum mechanics can be explored further by considering a certain variation of Feynman et al.'s second gedankenexperiment (Epstein, 1945; Renninger, 1960).[1] The results of this exploration are even more surprising than those presented by Feynman et al. in their gedanken-experiments. Empirical work on electron shelving that supports the next gedankenexperiment has been conducted by Nagourney, Sandberg, and Dehmelt (1986), Bergquist, Hulet, Itano, and Wineland (1986), and by Sauter, Neuhauser, Blatt, and Toschek (1986). This work has been summarized by Cook (1990).[2]

In a similar arrangement to that found in Gedankenexperiment 2, one can determine which of the two holes an electron went through on its way to the backstop by using a light that is placed near only one of the holes and which illuminates only the hole it is placed by (Figure 4). Illuminating only one of the holes yields a distribution of the electrons similar to that which one would expect if the light were placed between the holes, as in Feynman et al.'s second gedankenexperiment. The distribution is similar to the sum of the distributions of electrons that one would expect if only one or the other of the holes were open at a particular time.

---

[1] Epstein (1945) presented the essence of Gedankenexperiment 3 using the passage of photons through an interferometer. Renninger (1960) also discussed a gedankenexperiment in an article entitled "Observations without Disturbing the Object" in which the essence of Gedankenexperiment 3 is presented.

[2] In electron shelving, an ion is placed into a superposition of two quantum states. In each of these states, an electron of the ion is in one or the other of two energy levels. The transition to one of the quantum states occurs very quickly and the transition to the other state occurs very slowly. If the ion is repeatedly placed in the superposition of states after it transitions to one or the other of the superposed states, one finds the atomic electron in general transitions very frequently between the superposed quantum states and the quantum state characterized by the very quick transition. The photons emitted in these frequently occurring transitions to the quantum state characterized by the very quick transition are associated with resonance fluorescence of the ion. The absence of resonance fluorescence means that the ion has transitioned into the quantum state that occurs infrequently.

Cook (1990) has pointed out that in the work of Dehmelt and his colleagues on electron shelving involving the $Ba^+$ ion, the resonance fluorescence of a single ion is of sufficient intensity to be detectable by the dark-adapted eye alone, and the making of a negative observation, to be discussed shortly, is thus not dependent on any measuring device external to the observer.

Moreover, when an observer knows that electrons have passed through the unilluminated hole because they were not seen to pass through the illuminated hole, the distribution of these electrons through the unilluminated hole resembles the distribution of electrons passing through the illuminated hole (Figure 5). Consider also the point that if: 1) the light is turned off before sufficient time has passed allowing the observer to conclude that an electron could not have passed through the illuminated hole, and 2) an electron has not been observed at the illuminated hole, the distribution of many such electrons passing through the wall is determined by an interference pattern that is the sum of diffraction patterns of the waves of the electrons passing through the two holes similar to that found in Gedankenexperiment 1 (Epstein, 1945; Renninger, 1960).

## DISCUSSION OF THE GEDANKENEXPERIMENTS

The immediate question is how are the results in Gedankenexperiment 3 possible given Feynman et al.'s thesis that physical interaction between the light source and electron is necessary to destroy the interference? Where the light illuminates only hole A, electrons passing through hole B do not interact with photons from the light source and yet interference is destroyed in the same manner as if the light source illuminated both holes A and B. In addition, the distribution of electrons passing through hole B at the backstop indicates that there has been a change in the description of these electrons, even though no physical interaction has occurred between these electrons and photons from the light source.

Epstein (1945) maintained that these kinds of different effects on the physical world in quantum mechanics that cannot be ascribed to physical causes are associated with "*mental certainty*" (p. 134) on the part of an observer as to which of the possible alternatives for a physical existent occurs. Indeed, the factor responsible for the change in the wave function for an electron headed for holes A and B, and which is not illuminated at hole A, is *knowledge* by the observer as to whether there is sufficient time for an electron to pass through the "illuminated" hole. To borrow a term used by Renninger (1960), when the time has elapsed in which the electron could be illuminated at hole A, and it is not illuminated, the observer makes a "negative" (p. 418) observation.

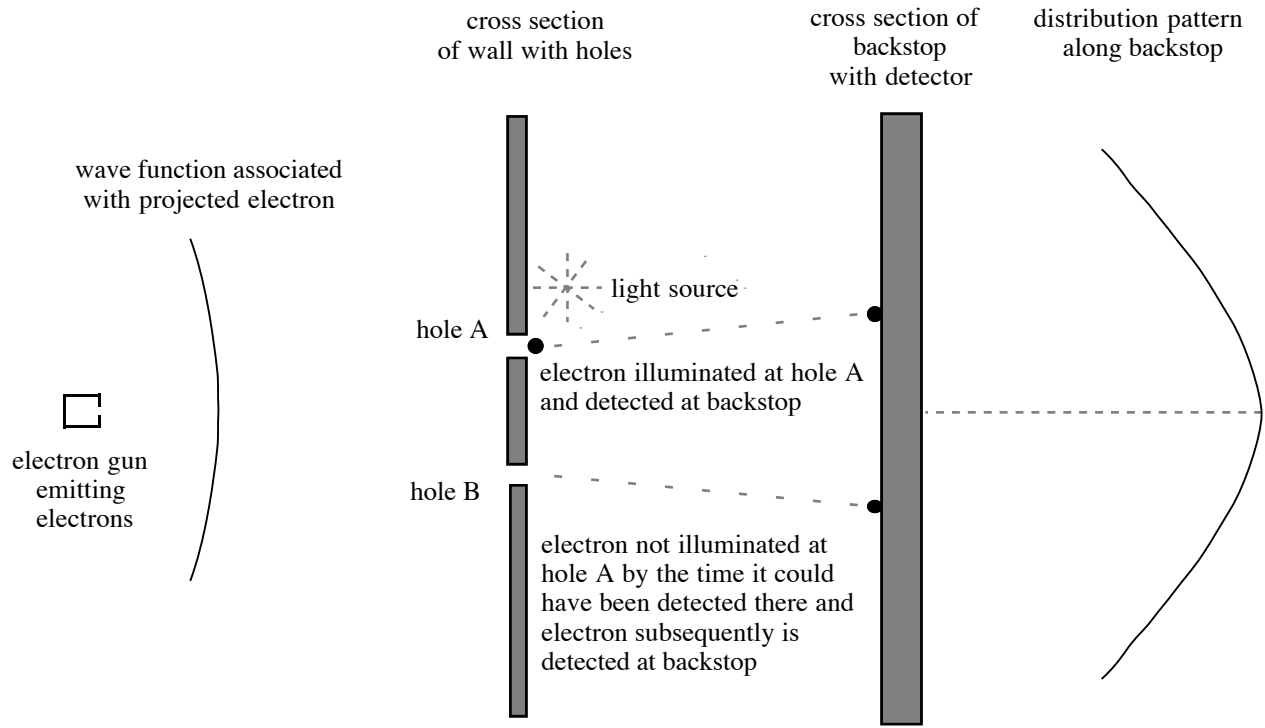

Figure 4

Two-hole gedankenexperiment with strong light source illuminating only one hole.
(Gedankenexperiment 3)

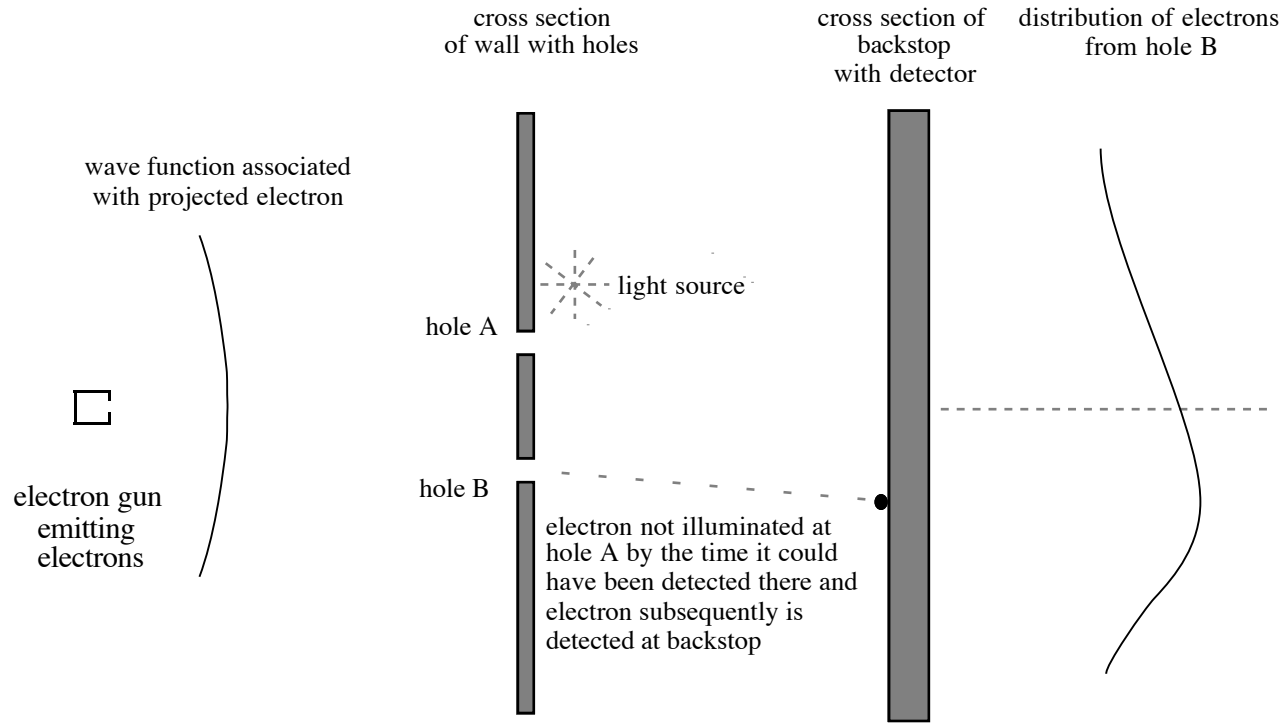

Figure 5

Two-hole gedankenexperiment with strong light source illuminating only one hole in which the distribution of electrons from unilluminated hole is shown.

# On the Nature

The common factor associated with the electron's passage through the wall in a manner resembling that found for classical-like particles in Gedankenexperiments 2 and 3 is the observing, thinking individual's knowledge as to whether an electron passed through a particular hole. The physical interaction between photons from the light source and electrons passing through either hole 1 or hole 2 is not a common factor. It should be remembered that some causative factor is implied by the very different electron distributions in Gedankenexperiments 1 and 2. It is reasonable to conclude that knowledge by the observer regarding the particular path of the electron through the wall is a factor in the change in the distribution of the electrons in Gedankenexperiment 1 to that found for electrons in Gedankenexperiments 2 and 3.

It might be argued that in Gedankenexperiment 3 a non-human recording instrument might record whether or not an electron passed through the illuminated hole in the time allowed, apparently obviating the need for a human observer. But, as has been shown, a non-human recording instrument is not necessary to obtain the results in Gedankenexperiment 3. And yet even if a non-human instrument is used, ultimately a person is involved to read the results who could still be responsible for the obtained results. Furthermore, one would still have to explain the destruction of the interference affecting the distribution of the electrons at the backstop without relying on a physical interaction between the electrons and some other physical existent. Without ultimately relying on a human observer, this would be difficult to accomplish when the non-human recording instrument presumably relies on physical interactions for its functioning.

It should also be emphasized that the change in the wave function for an electron passing through the unilluminated hole in Gedankenexperiment 3 provides the general case concerning what is necessary for the change in a wave function to occur in a measurement of the physical existent with which it is associated. It was shown clearly in the extension of Feynman et al.'s gedankenexperiments that the change in the wave function of an electron or other physical existent is not due fundamentally to a physical cause. Instead, the change in the wave function is linked to the knowledge attained by the observer of the circumstances affecting the physical existent measured.

# On the Nature